\let\JAPKernelLabel\label
\documentclass[
    aip,
    jap,
    reprint,
    floatfix,
    amsmath,
    amssymb,
    superscriptaddress
]{revtex4-2}

\usepackage{graphicx}
\usepackage{bm}
\usepackage{siunitx}
\usepackage{xcolor}
\makeatletter
\IfFormatAtLeastTF{2023-06-01}{%
  \let\label\JAPKernelLabel
  \let\ltx@label\JAPKernelLabel
}{}
\makeatother
\usepackage{hyperref}
\hypersetup{hidelinks}


\graphicspath{{figures/}{./}}

\newcommand{\ket}[1]{|#1\rangle}
\newcommand{\bra}[1]{\langle #1|}

\newcommand{\includeFigureOrPlaceholder}[2]{%
\IfFileExists{figures/#1}{%
\includegraphics[#2]{figures/#1}%
}{%
\IfFileExists{#1}{%
\includegraphics[#2]{#1}%
}{%
\fbox{%
\begin{minipage}[c][0.22\textheight][c]{0.88\linewidth}
\centering
Missing figure file: \texttt{\detokenize{#1}}\\
Add the file to \texttt{figures/} or the manuscript directory.
\end{minipage}}%
}%
}%
}

\begin{document}

\title{Time- and phase-resolved emulation of sign-selective counterdiabatic dynamics in a superconducting qutrit}

\author{Yiwen Li}
\thanks{These authors contributed equally to this work.}
\affiliation{National Laboratory of Solid State Microstructures, School of Physics, Nanjing University, Nanjing 210093, China}
\affiliation{Shishan Laboratory, Suzhou Campus of Nanjing University, Suzhou 215000, China}
\affiliation{Jiangsu Key Laboratory of Quantum Information Science and Technology, Nanjing University, Suzhou 215163, China}
\affiliation{Synergetic Innovation Center of Quantum Information and Quantum Physics, University of Science and Technology of China, Hefei, Anhui 230026, China}
\affiliation{Hefei National Laboratory, Hefei 230088, China}

\author{Luojia Xu}
\thanks{These authors contributed equally to this work.}
\affiliation{Beijing Key Laboratory of Fault-Tolerant Quantum Computing, Beijing Academy of Quantum Information Sciences, Beijing 100193, China}
\affiliation{Beijing National Laboratory for Condensed Matter Physics, Institute of Physics, Chinese Academy of Sciences, Beijing 100190, China}
\affiliation{University of Chinese Academy of Sciences, Beijing 100049, China}

\author{Xinsheng Tan}
\email{tanxs@nju.edu.cn}
\affiliation{National Laboratory of Solid State Microstructures, School of Physics, Nanjing University, Nanjing 210093, China}
\affiliation{Shishan Laboratory, Suzhou Campus of Nanjing University, Suzhou 215000, China}
\affiliation{Jiangsu Key Laboratory of Quantum Information Science and Technology, Nanjing University, Suzhou 215163, China}
\affiliation{Synergetic Innovation Center of Quantum Information and Quantum Physics, University of Science and Technology of China, Hefei, Anhui 230026, China}
\affiliation{Hefei National Laboratory, Hefei 230088, China}

\author{Yang Yu}
\email{yuyang@nju.edu.cn}
\affiliation{National Laboratory of Solid State Microstructures, School of Physics, Nanjing University, Nanjing 210093, China}
\affiliation{Shishan Laboratory, Suzhou Campus of Nanjing University, Suzhou 215000, China}
\affiliation{Jiangsu Key Laboratory of Quantum Information Science and Technology, Nanjing University, Suzhou 215163, China}
\affiliation{Synergetic Innovation Center of Quantum Information and Quantum Physics, University of Science and Technology of China, Hefei, Anhui 230026, China}
\affiliation{Hefei National Laboratory, Hefei 230088, China}

\date{\today}

\begin{abstract}
In cyclic three-level systems, phase-sensitive interference can convert a coupling-sign reversal into a population contrast, providing a controllable analogue of handedness. We emulate this mechanism in a superconducting transmon qutrit using two resonant single-photon drives and a phase-controlled effective two-photon drive. Effective loop phases \(\Phi_L=+\pi/2\) and \(\Phi_R=-\pi/2\), separated by \(\pi\), reverse the auxiliary contribution to the dark--bright coupling, canceling the intrinsic nonadiabatic term for \(L\) and reinforcing it for \(R\). Time-resolved trajectories reveal dark-pathway transfer for \(L\) and bright-manifold participation for \(R\). Measured late-time averages give a target-state population contrast of 0.929. Loop-phase measurements establish phase selectivity, while pulse-area and common-mode amplitude scans characterize the operating regime and control sensitivity. These results provide direct time- and phase-domain access to sign-selective counterdiabatic interference in a programmable cyclic qutrit.
\end{abstract}

\maketitle

\section{Introduction}

\begin{figure*}[!t]
\centering
\includeFigureOrPlaceholder{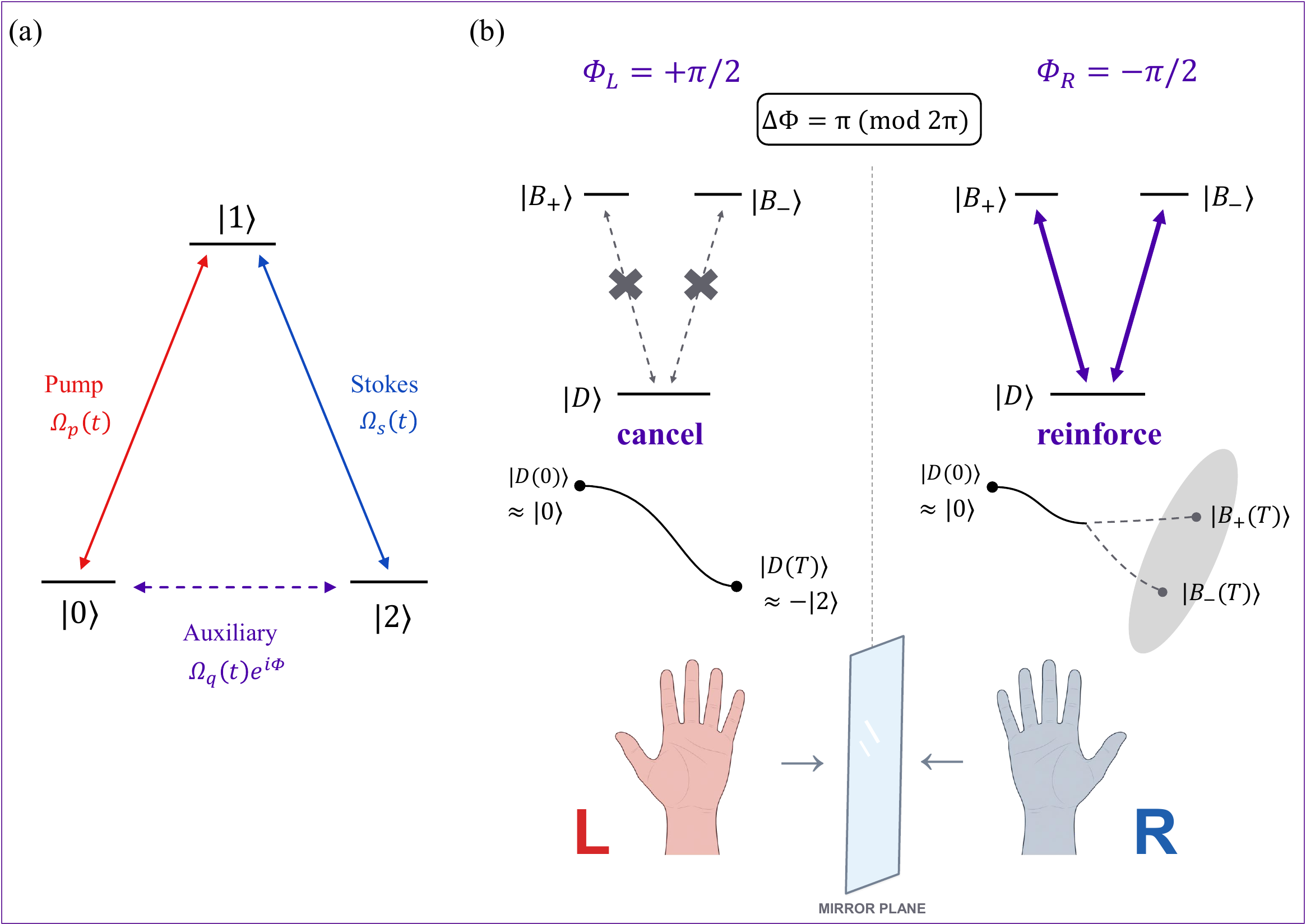}{width=\textwidth}
\caption{
\textbf{Phase-selective counterdiabatic interference in a cyclic qutrit.}
(a) Pump and Stokes couplings connect adjacent transitions, while a phase-controlled effective two-photon coupling closes the cyclic loop through \(\ket{0}\leftrightarrow\ket{2}\).
(b) Schematic of the interference mechanism in the moving adiabatic basis. In the \(L\) branch, \(\Phi_L=+\pi/2\), the auxiliary counterdiabatic contribution cancels the intrinsic nonadiabatic dark--bright coupling, allowing the state to follow the dark pathway. Shifting the effective loop phase by \(\pi\) gives the \(R\) branch, \(\Phi_R=-\pi/2\), and reverses the auxiliary contribution. The two contributions then reinforce, producing bright-manifold participation. Here, \(L\) and \(R\) denote the two programmable synthetic-handedness branches.
}
\label{fig:concept}
\end{figure*}

The dynamics of a cyclic three-level system depend on the gauge-invariant phase accumulated around its coupling loop. In chiral molecules, the product of three transition dipole moments changes sign between enantiomers, producing a \(\pi\) loop-phase difference~\cite{Kral2001,LiBruder2008}. Microwave experiments have exploited this sign change for chiral analysis, enrichment, and state transfer~\cite{Patterson2013Nature,PattersonDoyle2013PRL,Shubert2014,Eibenberger2017,Perez2017,Leibscher2019,Sun2023,Lee2024NatCommun}. Theory has further clarified the roles of cyclic manifolds, degeneracy, interference, and pulse-area control~\cite{Ye2018,Leibscher2022,Wu2020,Guo2022}. Together, these studies establish loop phase as a resource for mapping molecular handedness onto state populations. However, an endpoint contrast does not reveal how the sign reversal redirects the dynamical pathway. It also does not identify how finite pulse area determines whether that pathway remains discriminating.

Counterdiabatic control provides a direct mechanism that links this phase reversal to the dynamics. Shortcuts to adiabaticity suppress nonadiabatic transitions through auxiliary control~\cite{DemirplakRice2003,Berry2009,Chen2010,GueryOdelin2019}. They have been demonstrated in atomic, spin, and solid-state systems~\cite{Bason2012,Zhang2013,Du2016,Zhou2017}. In superconducting circuit QED, invariant-based shortcuts have also been proposed to accelerate the preparation of entangled states between two qubits~\cite{YanFeng2021}. In stimulated Raman adiabatic passage (STIRAP), an auxiliary counterdiabatic coupling between the initial and target states closes the loop~\cite{Giannelli2014,VitanovRMP2017}. Dressed-state control provides an alternative approach~\cite{Baksic2016}. The loop phase determines whether the auxiliary contribution cancels or reinforces the intrinsic nonadiabatic dark--bright coupling~\cite{VitanovDrewsen2019}. A finite pump--Stokes area then converts this pathway difference into a large endpoint contrast. Verifying the mechanism therefore requires time-resolved trajectories together with continuous phase and area measurements.

Testing this mechanism requires programmable coupling phases and level-resolved readout. Transmon qutrits provide both capabilities~\cite{Koch2007,Bianchetti2010,Peterer2015,Morvan2021}. Superconducting experiments have established coherent population trapping and electromagnetically induced transparency~\cite{Kelly2010,Abdumalikov2010}. STIRAP has been realized in ladder-type superconducting qutrits~\cite{XuNatCommun2016,Kumar2016} and extended to coherent state transfer between superconducting qubits mediated by a tunable coupler~\cite{LiAPL2021}. Counterdiabatic protocols enable accelerated population transfer~\cite{VepsalainenQST2018,Vepsalainen2019}, while optimized shortcut protocols reshape the STIRAP pulses to accelerate transfer and suppress leakage in weakly anharmonic superconducting qudits~\cite{Zheng2022}. Related experiments have demonstrated holonomic control and phase-dependent dynamics~\cite{Danilin2018,Tao2021}, including holonomic synthetic chiral discrimination with time-dependent population mapping~\cite{XuAPL2024}. Building on these advances, the present study resolves how a phase-only reversal under identical pulse envelopes switches between counterdiabatic cancellation and reinforcement, and how loop phase and finite pulse area determine the resulting population contrast.

Here, we implement this test in a superconducting transmon qutrit. The pump, Stokes, and auxiliary envelopes are identical for the two programmed synthetic-handedness branches. Their effective loop phases differ by \(\pi\), which isolates the coupling-sign reversal. Time-resolved reconstruction of all three populations separates dark-pathway transfer from bright-manifold dynamics and reveals a large target-state contrast. Continuous phase and finite-area scans distinguish the roles of these two control parameters. A common-mode amplitude scan then probes the aggregate response to calibration and waveform errors.

\begin{figure*}[!t]
\centering
\includeFigureOrPlaceholder{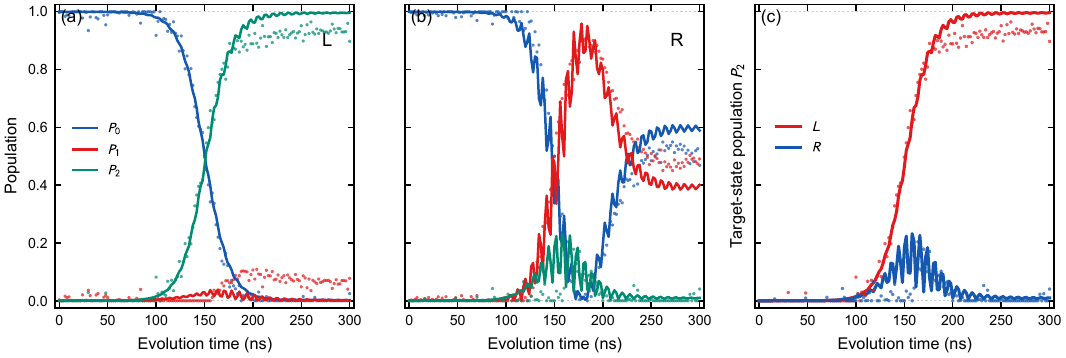}{width=\textwidth}
\caption{
\textbf{Time-resolved dynamics.}
(a,b) Populations for the \(L\) and \(R\) branches, respectively. (c) Direct comparison of their target-state populations. The envelopes are identical, and the branches differ only by a \(\pi\) effective-loop-phase shift. Symbols show measurements at 151 truncation times, and solid lines show laboratory-frame calculations.
}
\label{fig:dynamics}
\end{figure*}

\section{Phase-selective shortcut mechanism}

Retaining the phases of all three drives, the resonant rotating-frame Hamiltonian in the ordered basis \(\{\ket{0},\ket{1},\ket{2}\}\) is

\begingroup
\setlength{\arraycolsep}{1.2pt}
\begin{equation}
H(t)
=
\frac{\hbar}{2}
\begin{pmatrix}
0 & \Omega_p(t)e^{i\varphi_p} & \Omega_q(t)e^{i(2\varphi_q+\pi)} \\
\Omega_p(t)e^{-i\varphi_p} & 0 & \Omega_s(t)e^{i\varphi_s} \\
\Omega_q(t)e^{-i(2\varphi_q+\pi)} & \Omega_s(t)e^{-i\varphi_s} & 0
\end{pmatrix}.
\label{eq:H}
\end{equation}
\endgroup
Here, \(\Omega_{p,s,q}\) are real envelopes, \(\varphi_p\) and \(\varphi_s\) are the pump and Stokes phases, and \(\varphi_q\) is the phase of the physical two-photon drive. Taking the pump and Stokes phases as references reduces Eq.~\eqref{eq:H} to the gauge-fixed form \(H(\Phi,t)\), where \(\Phi=2\varphi_q-\varphi_p-\varphi_s+\pi\pmod{2\pi}\) is the effective loop phase. We define the \(L\) branch by \(\Phi_L=+\pi/2\) and the \(R\) branch by \(\Phi_R=-\pi/2\). The two settings use identical envelopes and differ only by a \(\pi\) effective-loop-phase shift. This shift reverses the auxiliary coupling sign (Fig.~\ref{fig:concept}). In our implementation, the \(L/R\) notation denotes programmable synthetic handedness. In the molecular analogue, the product of three transition dipole moments has opposite signs for the two enantiomers.

The physical consequence of this sign reversal is most transparent in the moving adiabatic basis. The pump and Stokes pulses define \(\tan\theta=\Omega_p/\Omega_s\) and \(\ket{D}=\cos\theta\ket{0}-\sin\theta\ket{2}\). The corresponding bright states are \(\ket{B_\pm}=(\sin\theta\ket{0}\pm\ket{1}+\cos\theta\ket{2})/\sqrt{2}\). The matrix element \(\mathcal C_\pm=\bra{B_\pm}(H/\hbar-i\partial_t)\ket{D}\) contains the intrinsic nonadiabatic term and the auxiliary contribution. At the two operating phases,
\begin{equation}
\begin{aligned}
\mathcal{C}_{\pm}^{(L)}(t)
&=\frac{i}{\sqrt{2}}\left[\dot{\theta}(t)-\frac{\Omega_q(t)}{2}\right],\\
\mathcal{C}_{\pm}^{(R)}(t)
&=\frac{i}{\sqrt{2}}\left[\dot{\theta}(t)+\frac{\Omega_q(t)}{2}\right].
\end{aligned}
\label{eq:interference}
\end{equation}
For \(\Omega_q=2\dot\theta\), the two terms in Eq.~(\ref{eq:interference}) cancel for the \(L\) branch. In contrast, Eq.~(\ref{eq:interference}) yields \(\mathcal{C}_{\pm}^{(R)}=i\sqrt{2}\dot\theta\) for the \(R\) branch. The \(L\) branch therefore follows the dark pathway, whereas the \(R\) branch enters the bright manifold. This phase-controlled pathway separation alone does not ensure endpoint discrimination. Suppressing the final \(R\)-branch population also requires a suitable finite pulse area. For the equal-width Gaussian Rabi envelopes used here, the pulse centers are separated by one width parameter \(\tau\), with the Stokes pulse preceding the pump pulse. We define their nominal pulse areas by
\begin{equation}
\begin{aligned}
\mathcal A_j &\equiv \int_{-\infty}^{\infty}\Omega_j(t)\,\mathrm{d}t \\
&=W_0\tau\sqrt{\pi}=W_{00}\pi,\qquad j=p,s,
\end{aligned}
\label{eq:pulse-area}
\end{equation}
where \(\Omega_j\) is an angular Rabi frequency and \(W_0=W_{00}\sqrt{\pi}/\tau\). The choice \(W_{00}=1.234\) therefore corresponds to a nominal area of \(1.234\pi\) for each pulse, near the ideal \(R\)-branch minimum for this Gaussian pair and the auxiliary prescription \(\Omega_q=2\dot\theta\)~\cite{VitanovDrewsen2019}. It specifies an operating point for these pulse conditions; changing the shape or relative delay generally changes the area at which the minimum occurs. At substantially larger areas, both branches approach adiabatic transfer, and the handedness contrast diminishes.

\section{Experimental implementation}

\begin{figure*}[!t]
\centering
\includeFigureOrPlaceholder{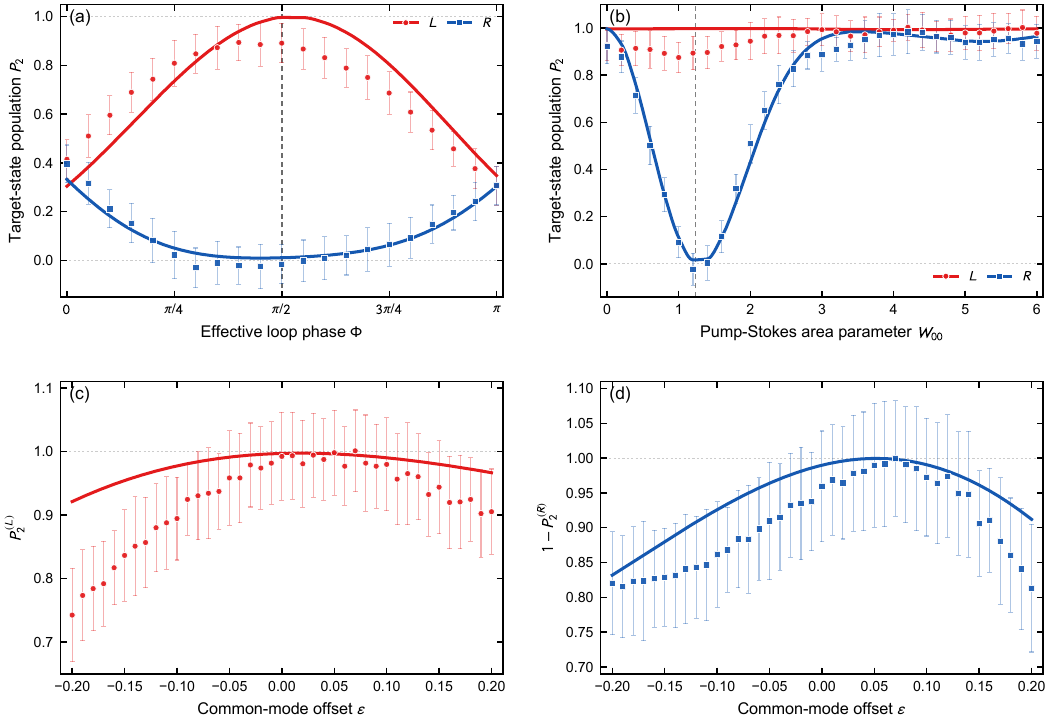}{width=\textwidth}
\caption{
\textbf{Effective loop-phase dependence, finite-area response, and common-mode amplitude response.}
(a) Target-state population versus effective loop phase. The \(L\) branch is shown for \(\Phi\in[0,\pi]\), while the \(R\) branch corresponds to the phase-shifted setting \(\Phi+\pi\). (b) Target-state population versus the pump--Stokes area parameter \(W_{00}\). The dashed line marks the nominal value of 1.234. (c,d) \(P_2^{(L)}\) and \(1-P_2^{(R)}\) under \(A_{p,s}\!\to\!(1+\epsilon)A_{p,s}\) and \(A_q\!\to\!\sqrt{1+\epsilon}A_q\). All populations in (a)--(d) are evaluated at the end of the \(T_{\mathrm g}=\SI{300}{\nano\second}\) sequence. Symbols show measurements, error bars denote one standard deviation, and solid lines show calculations.
}
\label{fig:phase-area}
\end{figure*}

We realize the cyclic Hamiltonian in the three lowest levels of a transmon. The measured transition frequencies are \(\omega_{01}/2\pi=\SI{5.242}{\giga\hertz}\) and \(\omega_{12}/2\pi=\SI{4.988}{\giga\hertz}\). They correspond to an anharmonicity of \(\alpha/2\pi=-\SI{254}{\mega\hertz}\). The relaxation times are \(T_1^{01}=\SI{31.4}{\micro\second}\) and \(T_1^{12}=\SI{11.3}{\micro\second}\). The Ramsey times are \(T_{2,\mathrm R}^{01}=\SI{24.5}{\micro\second}\), \(T_{2,\mathrm R}^{12}=\SI{9.75}{\micro\second}\), and \(T_{2,\mathrm R}^{02}=\SI{9.13}{\micro\second}\).

Each sequence begins with the qutrit in \(\ket{0}\) and lasts \(T_{\mathrm g}=\SI{300}{\nano\second}\). We set \(t_c=T_{\mathrm g}/2\) and \(\tau=T_{\mathrm g}/2\pi\). The physical pump and Stokes envelopes are
\[
\begin{aligned}
A_p(t)
&=W_0\exp\!\left[-\frac{(t-t_c-\tau/2)^2}{\tau^2}\right],\\
A_s(t)
&=\frac{W_0}{\sqrt{2}}
  \exp\!\left[-\frac{(t-t_c+\tau/2)^2}{\tau^2}\right].
\end{aligned}
\]
With the calibrated mappings \(\Omega_p=A_p\) and \(\Omega_s=\sqrt{2}A_s\), these choices give equal-peak Gaussian Rabi envelopes centered at \(t_c+\tau/2\) and \(t_c-\tau/2\), respectively. The Stokes pulse therefore precedes the pump pulse by \(\tau\). We use \(W_0=1.234\sqrt{\pi}/\tau\), corresponding to \(W_0/2\pi\simeq\SI{7.29}{\mega\hertz}\). The nominal areas in Eq.~\eqref{eq:pulse-area} refer to the full Gaussian envelopes; truncation to \(0\leq t\leq T_{\mathrm g}\) changes each area by less than \(0.01\%\). The two-photon branch is driven at \(\omega_q/2\pi=\SI{5.115}{\giga\hertz}\) with
\[
A_q(t)=2^{1/4}\sqrt{\frac{|\alpha|}{\tau}}\,
\operatorname{sech}^{1/2}\!\left[\frac{2(t-t_c)}{\tau}\right],
\]
whose peak is \(A_{q,\max}/2\pi\simeq\SI{34.6}{\mega\hertz}\). The Gaussian pair gives \(\tan\theta=\Omega_p/\Omega_s=\exp[2(t-t_c)/\tau]\) and \(\dot\theta=\tau^{-1}\operatorname{sech}[2(t-t_c)/\tau]\). To leading order, \(\Omega_q\simeq\sqrt{2}A_q^2/|\alpha|=2\tau^{-1}\operatorname{sech}[2(t-t_c)/\tau]=2\dot\theta\), as required for the cancellation branch.

We compensate the leading dynamic ac Stark shift of the strong two-photon tone using \(\dot\chi=-A_q^2/(2|\alpha|)\)~\cite{VepsalainenQST2018}. The physical two-photon phases of the \(L\) and \(R\) sequences differ by \(\pi/2\). Phase doubling converts this difference into the required \(\pi\) effective-loop-phase reversal. The two sequences therefore share the same durations and envelopes and differ only in the programmed auxiliary phase.

The solid curves are obtained from decoherence-free laboratory-frame calculations of a three-level system. The model retains the physical carriers, counter-rotating terms, off-resonant couplings, implemented envelopes, and phase modulation. A calibrated readout-response matrix reconstructs the populations. Time-resolved population triplets are projected onto the probability simplex. We omit error bars because this scan was not independently repeated. The phase, area, and amplitude scans use linear correction without clipping, and their error bars denote one standard deviation.

\section{Handedness discrimination in the time domain}

To resolve the pathways underlying the endpoint contrast, we terminate all three drives at 151 evolution times and reconstruct the populations. The \(L\) branch transfers population from \(\ket{0}\) to \(\ket{2}\) with only weak transient \(\ket{1}\) occupation [Fig.~\ref{fig:dynamics}(a)]. Over \(\SI{240}{\nano\second}\leq t\leq\SI{300}{\nano\second}\), the measured and calculated means of \(P_2^{(L)}\) are 0.932 and 0.995, respectively. The measured trajectory is therefore consistent with the cancellation branch predicted by Eq.~(\ref{eq:interference}).

Relaxation and dephasing during the \(\SI{300}{\nano\second}\) sequence can reduce the measured \(L\)-branch population relative to the decoherence-free calculation. Amplitude-calibration errors, relative-phase errors, and waveform distortion can perturb the dark--bright cancellation, while imperfect dynamic Stark compensation can introduce residual detuning. Readout-correction uncertainty can also affect the reconstructed populations.

Changing only the physical auxiliary phase by \(\pi/2\), equivalent to a \(\pi\) effective-loop-phase shift, produces pronounced transient \(\ket{1}\) occupation. It also suppresses transfer in the \(R\) branch [Fig.~\ref{fig:dynamics}(b)]. The measured and calculated late-time means of \(P_2^{(R)}\) are 0.003 and 0.011, respectively. These values yield an experimental handedness contrast of \(\Delta P_2=P_2^{(L)}-P_2^{(R)}=0.929\). Because the two sequences have identical envelopes, the contrast originates from the coupling-sign reversal rather than a pulse-energy difference. The complete trajectories show that endpoint discrimination arises from dark-pathway transfer for \(L\) and coherent bright-manifold participation for \(R\).

\section{Loop-phase dependence, finite-area response, and amplitude errors}

To connect the two nominal branches through a single control parameter, we continuously vary the auxiliary phase. The target-state populations change smoothly [Fig.~\ref{fig:phase-area}(a)] and reach their largest separation near \(\Phi=\pi/2\). Residual ac Stark detuning, off-resonant couplings, and counter-rotating terms slightly shift the calculated \(R\)-branch minimum. The measurements reproduce the overall phase dependence. This continuous interpolation places \(L\) and \(R\) at opposite points of one coherent loop-phase response. It also links the handedness contrast directly to Eq.~(\ref{eq:H}).

The phase reversal becomes strongly discriminating only near a selected finite-area operating point. With the auxiliary pulse fixed, we scan \(W_{00}\) from 0 to approximately 6 using \(W_0=W_{00}\sqrt{\pi}/\tau\). Near the nominal value of 1.234, the \(L\) branch remains close to complete transfer, whereas the \(R\) branch reaches its minimum [Fig.~\ref{fig:phase-area}(b)]. At larger areas, \(P_2^{(R)}\) recovers as both branches approach adiabatic transfer~\cite{VitanovDrewsen2019}. The two scans therefore reveal complementary roles. Loop phase selects cancellation or reinforcement, whereas finite area converts the pathway difference into endpoint discrimination.

We finally apply a common-mode amplitude offset \(\epsilon\in[-0.2,0.2]\), following the common Rabi-amplitude error scan used in related transmon control experiments~\cite{XuAPL2024}. We scale \(A_{p,s}\) by \(1+\epsilon\) and the physical two-photon amplitude \(A_q\) by \(\sqrt{1+\epsilon}\). The latter scaling follows from \(\Omega_q\propto A_q^2\), so all nominal effective Rabi frequencies scale by \(1+\epsilon\) [Fig.~\ref{fig:phase-area}(c,d)]. The central discrimination feature persists around the calibrated point. However, the measured response is asymmetric about \(\epsilon=0\) and deviates more strongly from the idealized calculation near the scan edges. In the experiment, an amplitude offset affects more than just the integrated pulse area. The three tones are delivered at different carrier frequencies and power levels. The nonideal transfer functions of the waveform generator, mixers, and signal lines can therefore introduce tone-dependent amplitude and temporal distortions under nominal common-mode scaling. Moreover, both \(\Omega_q\) and the dynamic Stark shift are proportional to \(A_q^2\). Gain compression or bandwidth distortion can thus perturb the counterdiabatic coupling and residual detuning simultaneously. Because the nominal phase correction remains fixed during the scan, these changes can also introduce an \(\epsilon\)-dependent phase error. These mechanisms, together with decoherence and readout uncertainty, may jointly shape the measured asymmetry. The scan characterizes this aggregate response along the tested common-mode direction.

\section{Discussion and conclusion}

The measurements establish two connected aspects of phase-selective handedness resolution. In the time domain, a phase-only sign reversal separates dark-pathway transfer from bright-manifold dynamics. This separation produces a late-time contrast of 0.929. The complementary control scans place both branches on one coherent phase response. They also show that endpoint discrimination requires a selected finite-area operating point rather than an arbitrary pulse strength.

The common-mode offset scan probes the response to aggregate control errors around the calibrated operating point. The observed asymmetry is consistent with the combined sensitivities of the effective two-photon coupling, Stark compensation, and delivered waveforms. The decoherence-free laboratory-frame calculations reproduce the main trajectory, phase, area, and amplitude-response features and provide a common reference for identifying the residual deviations.

This experiment realizes the phase-selective cyclic mechanism proposed for chiral resolution~\cite{VitanovDrewsen2019} using programmable synthetic handedness. In molecules, the corresponding sign reversal arises from the enantiomer-dependent product of three transition dipole moments under common driving fields. Extending the protocol to molecular ensembles requires accounting for the initial rotational-state distribution, magnetic-sublevel degeneracy, and polarization-dependent transition strengths~\cite{Ye2018,Leibscher2019,Leibscher2022}. Molecular coherence times and the available control of pulse amplitudes and phases would also affect the attainable contrast. These time- and phase-resolved measurements isolate sign-selective counterdiabatic interference and complement holonomic population mapping in a transmon~\cite{XuAPL2024}.

\begin{acknowledgments}
This work was supported by the National Key R\&D Program of China (Grant Nos. 2022YFA1405304 and 2024YFA1409300), the National Natural Science Foundation of China (Grant Nos. 12504588, U21A20436, and 12074179), Quantum Science and Technology-National Science and Technology Major Project (Grant No. 2021ZD0301702), the Natural Science Foundation of Jiangsu Province (Grant Nos. BE2021015-1, BK20232002, and BK20233001), and the Natural Science Foundation of Shandong Province (Grant No. ZR2023LZH002).
\end{acknowledgments}

\section*{Author Declarations}

\subsection*{Conflict of Interest}
The authors have no conflicts to disclose.

\subsection*{Author Contributions}
Yiwen Li and Luojia Xu contributed equally to this work.
Yiwen Li: Conceptualization, Methodology, Investigation, Data curation,
Writing -- original draft.
Luojia Xu: Methodology, Validation, Formal analysis,
Writing -- review and editing.
Xinsheng Tan: Conceptualization, Supervision, Project administration,
Funding acquisition, Writing -- review and editing.
Yang Yu: Conceptualization, Supervision, Funding acquisition,
Writing -- review and editing.
All authors discussed the results and approved the final manuscript.

\section*{Data Availability}
The data that support the findings of this study are available from the
corresponding authors upon reasonable request.

\end{document}